# Modification of Tin (Sn) Metal Surfaces by Surface Plasmon Polariton Excitation

Vamsi Borra[1,3], Srikanth Itapu[2,3], Victor G. Karpov[4], Daniel G. Georgiev[3]

[1] Electrical and Computer Engineering, Youngstown State University, Youngstown, OH 44555, U.S.A.

[2] Department of ECE, Alliance College of Engineering & Design, Alliance University, Bengaluru, India 562106

[3] Department of EECS, University of Toledo, Toledo, OH 43606-3390, U.S.A.

[4] Department of Physics and Astronomy, University of Toledo, Toledo, OH 43606-3390, U.S.A.

**ABSTRACT**

We report on the modification of tin (Sn) film surfaces under a laser beam irradiation that triggers surface plasmon polariton (SPP) excitations. The observed surface features in the form of small raised grains, with well-defined rooting, look similar to tin whisker nodules. We attribute the appearance of those features to the field-induced nucleation caused by the SPP related strong electric field. Possible implications of our findings include accelerated-life testing for tin whisker growth related reliability as well as applications to nanoparticle nucleation.

Sn-based solders that are quintessential in micro-electronic packaging often develop electrically conductive hair-like structures on their surfaces, which are referred to as metal whiskers (MWs). Various whisker growth mechanisms on Sn coatings were discussed in a detailed recent review[1]. In addition, other metals and metal alloys, such as Pb, Ag, Cd, Zn, In, Al, Au, and Al[2,3] are also known to form whiskers. Various hypotheses for the growth mechanism have been proposed over the last several decades, ranging from the popular and long standing ones related stress relieving phenomena[3-6] to the most recent electrostatic theory[4–6]. Whisker growth mitigation techniques have been used and investigated in the past, and among them the use of small amount of Pb alloyed with the Sn solder, which substantially reduces the whisker growth[10], is the most important approach although it is still not well understood. However, relatively recent regulations and standards (such as RoHS[7]) require significant reduction or complete elimination the use of Pb. In this regard, researchers[8–10] proposed that a Ni sublayer between a Cu substrate and a Sn layer has potential for whiskers growth mitigation but such heterostructures were rendered unreliable due to the observed whiskers appearance on Ni-treated multilayer ceramic chip capacitors and connectors[11,12].

Electrostatic theory predictions motivated a novel approach to replace Ni sublayers with NiO, which was studied in detail recently[13]. Whiskers did not grow on the sample, with NiO sublayer sandwiched between Sn and Cu, even after 36 months, whereas whiskers of size and density that are usually expected are observed on the control sample (Sn thin film on Cu substrate without NiO sublayer). In the context of metal surface modifications, a recent study[14] reported whisker nucleation on a Sn film using locally applied electric field. The voltage between the conductive

AFM tip and the sample was used to generate the field. Minuscule whiskers were observed at the spots where the AFM tip was placed for a certain period. The growth direction of those whiskers matches the vertical orientation of the field. Similarly, Killefer et. al.[15] demonstrated a non-destructive gamma-ray irradiation method to accelerated growth of Sn MWs. Whisker densities and lengths increased at an acceleration factor of ~50 at the irradiated spots. The observed enhancement in MW development was attributed to the appearance gamma-ray induced charged defects and their localized electrostatic fields, affecting the whisker kinetics. In addition, in a microscopic study[16], the intricacies of whisker growth as to why MWs' diameters are in the micron range (surpassing the usual nano-sizes of nuclei in solids), how the diameters remain almost unaffected in the course of MW growth, the nature of MW diameter stochasticity, and the cause of the recognised striation structure of MW side surfaces were addressed in detail.

Concomitant to the SPP excitations is the effect of strong electric field enhancement[17,18] in the near surface region of a metal. That enhancement is responsible for multiple observed effects such as fluorescence, surface enhanced Raman scattering, and infrared absorption. The empirically and theoretically estimated enhancement coefficients approach several orders of magnitude.

This work is motivated by a hypothesis that the enhanced SPP electric field may be strong enough to cause structural modifications on a metal surface. Indeed assuming for example, a 'moderate' laser field of ~ 1 kV/cm and the enhancement coefficient of 100, the enhanced field will have strength of 0.1 MV/cm, which is sufficient to trigger phase transitions in some materials[19], create conductive pathways responsible for dielectric breakdowns[20], trigger metal whisker growth[5], and induce protrusions on liquid metal surfaces[21]. Hence, our hypothesis here is that for a low surface tension metals, such as Sn[22], maintaining SPP at the metal surface, will result in field-induced nucleation and development of grains with prevalent alignment that are likely to characterize the onset growth of whiskers.

Prism coupling techniques or near field probes can be used to generate SPPs on metal surfaces by providing momentum matching between the photons and the SPP excitations[23],[24],[25,26]. In particular, the Otto and Kretschmann methods are attenuated total reflection geometries that have been widely used, together with various modifications, for generating SPPs on smooth metal surfaces[23]. As mentioned in our previous work [27], Otto configuration is suitable for our setup. A standard prism coupling using attenuated total internal reflection in the Otto experimental setup is schematically represented in Fig. 1. Here, $\varepsilon_p$, $\varepsilon_d$ and $\varepsilon_m$ are the dielectric constants of prism, dielectric material, and metal film, respectively. Thickness of dielectric material and the metal film are $t_d$ and $t_m$.

In order to overcome the difficulty of *varying the incidence angle* when working with a triangular prism (Otto configuration), we used the setup of Fig. 2 where a UV grade fused silica plano-convex cylindrical lens (Thorlabs, n=1.460 at 588nm) serves as the prism (Fig. 2(a)). Thin-film samples, consisting of 500nm (±40nm) thick Sn films coated on glass substrates[28], are mounted on the lens' planar face with the metal film in contact with the flat glass surface. The intrinsic unevenness of the glass surface acts as a spacer with random air gaps serving as the dielectric material between the metal and the prism.

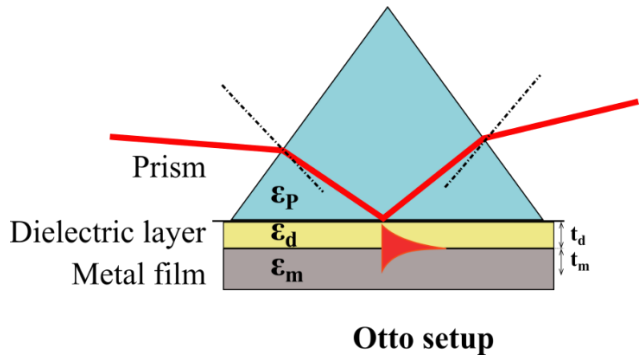

**Figure 1. The Otto prism coupling setup for exciting SPPs.**

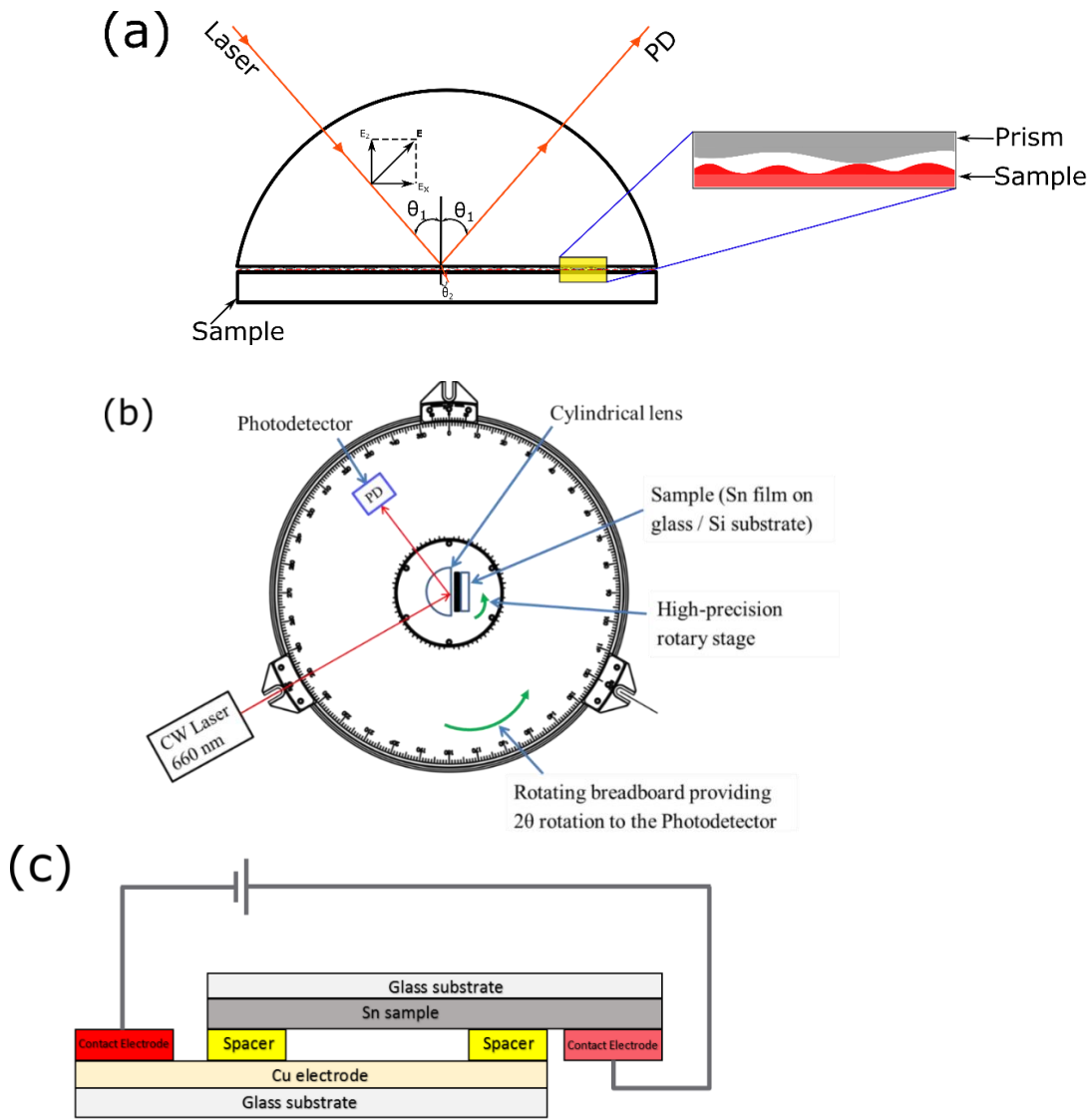

**Figure 2: (a) The Set-up for SPP generation, (b) sample mounting details, and (c) Parallel-plate capacitor setup.**

The experiment was conducted on a large-area rotating breadboard (Thorlabs) with the center section of the board replaced with a high precision rotary stage (model UTR, by Newport) (Fig. 2(b)). A 100mW CW semiconductor laser system (λ=662nm, model OBIS 660 by Coherent), mounted at a fixed location on an optical table, was used to excite the SPPs on the thin-film sample. A silicon photodiode (PD) based power probe was mounted on the rotating breadboard to record the reflected power from the sample's surface at various angles of incidence ($\theta_i$) and thus identify the matching angle for SPP generation.

The prism-sample combination was positioned on a precision rotary stage using a prism mount. The laser light is directed onto the sample (without focusing, so the laser spot diameter was about 1mm), starting at an angle of incidence (θ₁ on Fig. 2(a)) close to 90°. The high precision rotary stage (capable of step size of 0.017º) is aligned with the PD. The local minimum (dip) associated with SPP generation[23] was identified by plotting the reflected power vs. the angle of incidence.

We validated our SPP setup experimenting with silver (Ag) films. Ag has a small damping loss and SPPs are relatively easy to excite in it. We then conducted the same type of measurements on Sn films, which have high damping loss and in which SPPs are more difficult to excite.

The surface modification features in the laser/SPP treated areas were observed and identified by optical and scanning electron microscopy (SEM). Some of the samples were later subjected to additional electric fields (post-SSP) using a parallel-plate capacitor setup (Fig. 2(c)). The Sn sample was placed on a Cu coated glass electrode with a 50µm thick kapton tape used as a spacer. A voltage of 100V, was applied across this setup using a high voltage DC power supply (by Agilent). The electric field created in this configuration is estimated at $2\times10^4 \frac{V}{cm}$ . After this additional electric field treatment, the samples were again examined by optical microscopy and SEM.

The light intensity distribution at the metal-dielectric (the dielectric is air in this case) interface can be described with the Fresnel reflection and transmission coefficients. The corresponding equations for our setup were derived using equations 2 and 3 from Ref.[29]. They predict a deep minimum in the reflected laser intensity beyond the critical angle in the reflectance curve, which is attributed to the energy transfer from the optical wave to the SPP. It is reported based on theoretical[30],[23] and experimental[31] work that the angle ($\theta_p$) of incidence at which SPPs have maximum intensity is not exactly the same as the angle of incidence at which the reflectance exhibits a minimum. That deviation is due to the lag between the phase of the polarization field and the driving field [32]. In addition, this difference in ($\theta_p$) and ($\theta_i$) ranges in ±0.3° for noble metals (Ag, Au, etc.) due to their small damping losses. It can range anywhere between 25° - 35° for metals with large damping losses (Fe, Sn, Zn, etc.) [23,30].

For the experiment with Ag film, the deep minimum occurred at an incidence angle of 48.67°, as seen from Figure 3. This deviation from the theoretically predicted 45.09º could be ascribed to the trivial imprecision in prism mounting. In general, this observation confirms our experimental setup and shows that SPPs are generated on Ag film.

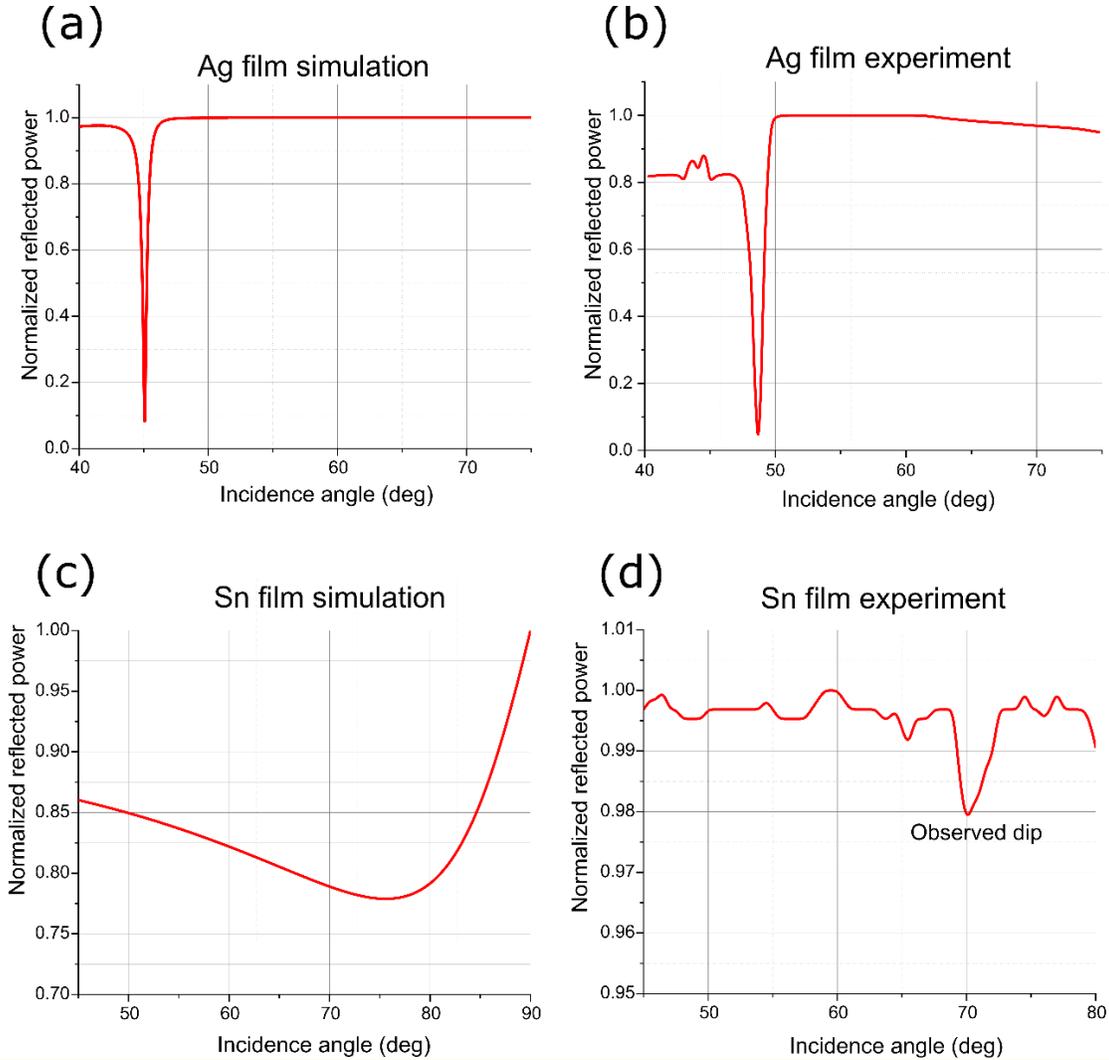

**Figure 3. Reflectance curves for Ag film ((a) simulation; (b) experiment) and Sn film ((c) simulation; (d) experiment).**

We repeated the above-mentioned simulation and an experimental run on the Sn (high damping) metal films. The dielectric function of Sn, at λ=662nm, is obtained from the literature[33]. Predictably, the SPP dip has a noticeable shift (Fig. 3 (c) and 3(d)) because of the larger damping. From the experimental plots we conclude that SPPs were generated at an incidence angle of 70° (see Fig. 3(d)). Similar shifts can also be detected with other metals that have notably large imaginary part ($\varepsilon_2$) in the dielectric function[23,30]. Note that the imaginary part of dielectric permittivity of Sn is approximately 13 times that of Ag.

The laser beam was left at the incidence angle at which the dip was observed for Sn films, at a power of 50mW for 5 hours, after detecting the reflectance dip. The film surface at and near the laser spot was then examined by SEM. Localized nuclei were observed at the irradiated spot as seen in Fig. 4, which was also reported earlier in[27]. On the other hand, no such surface features are observed away from the irradiated spot.

The sample was stored at room temperature and was periodically examined by SEM in order to register any further evolution. No significant changes were observed at the treated spot over the course of 7 weeks.

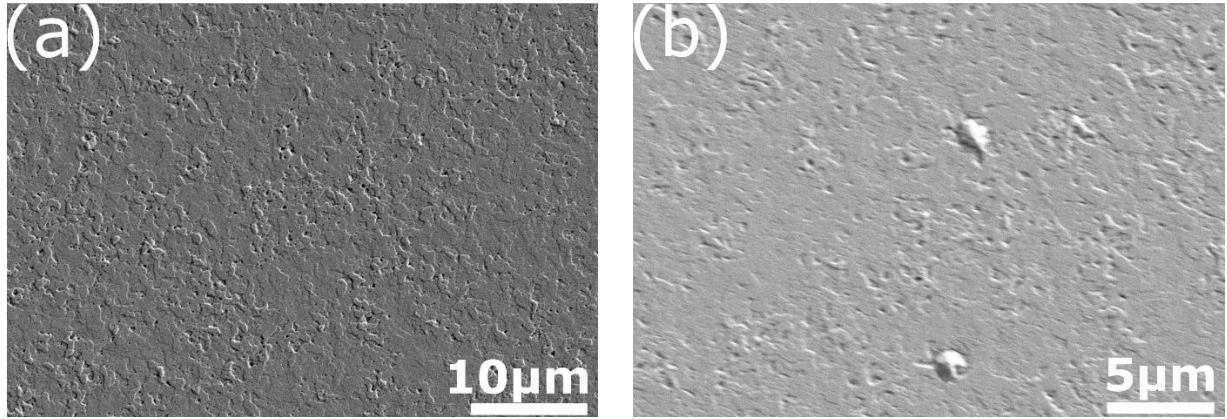

**Figure 4. (a) SEM scan away from the irradiated spot (No sign of nodules growth) and (b) at the irradiated spot after SPP experiment**

In an additional experiment, SPP treated, exactly the same way as described above, samples were later exposed to a constant electric field in the parallel plate capacitor setup of Fig. 2(c). After 12 hours of exposure to this field, some changes to the original nuclei (of the type in Fig. 4(b)) were observed. More specifically, as can be seen in Fig. 5, additional vertical growth of the original nuclei occurred. The vertical direction of growth is along the expected SPPs field direction. Fig. 5 (a), (b) and (c) show close-up SEM scans (top-view) of nuclei (Fig. 4(b)) after being treated with external electric field. The circular shallows (rooting) that are clearly visible in all three images suggest that material was transported towards the nuclei's base. Also, these top-view scans indicate that the diameter of some nuclei has increased by 80% compared to their original size (Fig 4(b)). Fig 5 (d) & (e) are lower magnification SEM images (top-view) of the nuclei after applying external electric field. A tilted SEM scan (Fig 5 (f)) on the same sample illustrates the vertical growth of the original nuclei.

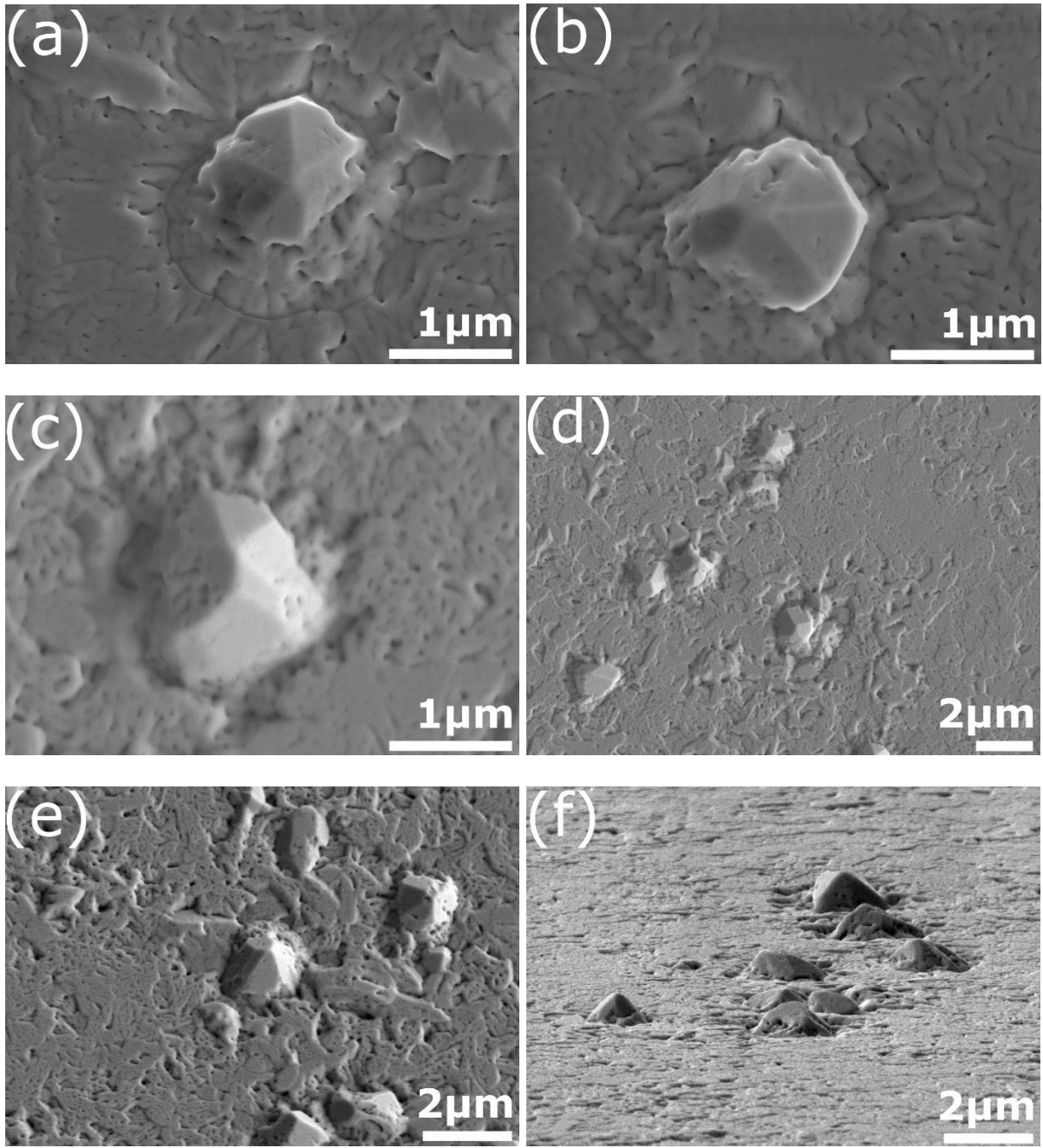

**Figure 5: (a) - (c) SEM images of the nuclei after applying an additional external E-field, showing the shallow circles around the base. (d) & (e) Lower-magnification SEM images of nuclei clusters. (f) Tilted SEM scan of the same sample showing the level of vertical growth.**

The above described nuclei exhibit features characteristic of metal whiskers (MW) and MW nodules[34], [35]. We recall that MW represent hair-like structures that often grow on metal surfaces, such as Sn, Zn. Short circuits in the sensitive equipment have caused substantial losses in the airspace, automotive, and other industries[36,37]. Researchers still don't come into consensus on the mechanism behind the MW growth. A few models attribute MW to stress relieving phenomena[38],[39],[40], while, other groups proposed[41],[3] different hypotheses; none of them leading to any quantitative estimates.

In particular, our observations are consistent with a consensus that whiskers grow from the root rather than the tip, which can be seen in Fig. 5 ((a) - (f)), evidenced by the observed root structure. Drifting of material towards the base of the nuclei indicates that the material required for its growth is supplied from relatively large distances through long-range, mostly lateral diffusion. However, unlike most reports on metal whiskers, we have seen few grooves or ridges on the outer surface[34], which is likely due to the limited vertical size, and the fact that the film is relatively thin and the inter-whisker distances are much smaller than in the case of regular whisker growth causing these nuclei (or whisker nuclei) to compete for material.

A recent electrostatic theory[4] provides some quantitative estimates of MW nucleation and growth rates, and their statistical distributions, which are consistent with experimental findings. The theory proposes that the small patches of net positive or negative electric charges formed due to the imperfections on metal surfaces can lead to the formation of the normal electric field $(E)$ which governs the whisker development. MW and nodule formation is explained as the electric-field-induced nucleation (FIN)[42] of needle shaped embryos. Their large induced dipole moments $p = \alpha E$ provide an energy gain $F_E = -p.E = \alpha E^2$ which outweighs the surface energy loss due to their nonspherical shape (here, $E$ is the electric field, and α ~ $h^3$ is the polarizability, where h is the embryo's length).

Note that the latter energy gain is independent of the sign of the electric field $(E)$, i. e. both the negatively and positively charged patches are equally efficient in whisker generation. Recent observations [5,43] confirmed the predicted effect of the electric fields on whisker growth.

Our observations are most consistent with the electrostatic theory[4] that in fact predicted possible effects SPP on whisker development. In particular, both the regions of negative and positive fields in SPP excitations can serve as efficient MW triggers. The SPP enhanced field strengths of the order of 0.1-1 MV/cm are in the ballpark of the theoretical predictions and field strengths used in independent experiments [5,6,44]. No other mechanisms would possibly explain the aforementioned observations.

The MW features observed here have rather limited lengths not exceeding 3 μm, which is much shorter than MW lengths of tens and hundreds on micron observed under electric field in the preceding work[5] . We note that the under-prism gap of ~ 1 μm available for whisker growth clearly limited the SPP caused MW growth. On the other hand, because the subsequent exposure to the capacitive DC field, provided enough space for MW development, we have to conclude that the external field of strength ~ 10 kV/cm is not a strong factor in the MW post-nucleation growth accelerating it rather moderately.

The latter conclusion is consistent with the electrostatic theory[4] showing that the nucleation rate is exponential in the field strength, i.e., $\nu \propto \exp[-\text{const}/(E_{DC} + E_{in})]$ where $E_{DC}$ and $E_{in}$ are respectively the external DC field and the intrinsic surface nonuniformity caused field. It is clear then that even for relatively small external field, $E_{DC} \ll E_{in}$, its effect can be rather significant as long as the exponent in nucleation range is large enough, which is typically the case. On the other hand, the growth rate was predicted to be quadratic in field strength, hence a relatively much lower from the external field component.

In summary, we (1) observed that surface plasmon polariton excitations can have strong enough impact on metal surfaces to induce their structural modifications in the form of metal whiskers and nodules, and (2) that observation is consistent with the electrostatic theory of metal whiskers. The above results shine light on the nature of metal whiskers and open a venue of metal surface patterning through SPP techniques. Also, they can have important practical implications: (i) leading to the development of accelerated failure testing methods of electronic components, which is currently not possible, or very difficult, due to the random and unpredictable nature of whiskers formation, and (ii) providing a technique for controllable formation of whiskers, which may lead to new fabrication methods of materials based on metals that are prone to whisker growth.

**ACKNOWLEDGEMENT**